\documentclass[aps,prd,preprint,groupedaddress,amsmath,amssymb,showpacs,showkeys]{revtex4-1}
\usepackage[utf8x]{inputenc}
\usepackage{picinpar}
\usepackage{epsfig}
\usepackage[english]{babel}
\usepackage{amsmath,amssymb,amsfonts,euscript}
\usepackage{bm}
\newcommand{\Deriv}[2]{\mbox{$\displaystyle \dfrac{{\rm D}#1}{{\rm D}#2}$}}
\newcommand{\kk}[1]{\,\mathrm{#1}}
\newcommand{\pot}[1]{\cdot 10^{#1}}
\newcommand{\w}{{\!}}

\newcommand{\Sp}{\;\! , \;\!}

\newcommand{\m}{m^{*}{}}
\newcommand{\fm}{(\mu)}
\newcommand{\fn}{(\nu)}
\newcommand{\Bm}[1]{\pmb{#1}}
\newcommand{\fa}{(\alpha)}
\newcommand{\fb}{(\beta)}

\newcommand{\fv}{(4)}
\newcommand{\fk}{(\kappa)}
\newcommand{\ft}{(\tau)}
\newcommand{\deriv}[2]{\,\mbox{$\displaystyle \dfrac{{\rm d}#1}{{\rm d}#2}$}\,}
\newcommand{\pderiv}[2]{\,\mbox{$\displaystyle\dfrac{{\partial}#1}{{\partial}#2}$}\,}

\DeclareMathOperator{\sign}{sign}
\DeclareMathOperator{\sn}{sn}
\DeclareMathOperator{\cn}{cn}
\DeclareMathOperator{\F}{F}

\begin{document}
\renewcommand{\figurename}{FIG}
\renewcommand{\tablename}{TABLE}

\title{Optical appearance of a compact binary system in the neighbourhood of supermassive black hole}

\author{A. Gorbatsievich }
\email[E-mail: ]{gorbatsievich@bsu.by}
\author{S. Komarov}
\email[E-mail: ]{staskomarov@tut.by}
\author{A. Tarasenko}
\email[E-mail: ]{tarasenk@tut.by}
\affiliation{Theoretical Physics Department, Belarusian State University, Nezavisimosti av., 4, 220030 Minsk, Belarus}
\thanks{A.G. acknowledges support from A.\,von Humboldt Foundation in the framework of the Institute Linkage Programm }

\date{\today}

\begin{abstract}
Optical appearance of a compact binary star in the field of a supermassive black hole is modeled in a strong field regime.  Expressions for the redshift, magnification coefficient and pulsar extinction time are derived.

By using the vierbein formalism we have derived equations of motion of compact binary star in the external gravitational field. We have analysed both the evolution of redshift of the optical ray from the usual star or white dwarf, and the times of arrival of pulses of pulsar. The results are illustrated by a calculation for a model binary system for the case of external gravitational field of a Schwarzschild black hole. The obtained results can be used for fitting timing data from the X-ray pulsars that moves in the neighbourhood of the Galactic Center (Sgr A*).
\end{abstract}


\maketitle

\renewcommand{\figurename}{FIG}
\renewcommand{\tablename}{TABLE}

\section{Introduction\label{Introduction}}
Since the discovery of binary pulsar B1913+16 by Hulse and Taylor in 1975 (see \cite{Taylor} for review) many new possibilities for testing theories of gravity have appeared. But most of these tests are performed only for the case of weak gravitational field. One of the challenging ways for the purpose of testing gravity in a strong field regime gives us the studying of the motion of the astrophysical objects near the supermassive black hole. The recent investigations that have been completed by cosmic observatories Chandra and XMM-Newton provides us evidences for existence supermassive black hole in the Galactic Center (Sgr A*) \cite{Baganoff, GBHL}.  Also it have provided evidences for plenty of binary pulsars and stars in this region \cite{X-ray, Stars}. In a volume of $1\ pc^{3}$ around SgrA*, there are $\sim 10^{4}$ compact objects of about one stellar mass \cite{MUN, Ayal};
presumably, about half
of these objects are bounded in binary systems (NS-NS, NS-BH and BH-BH).

Therefore it is possible to perform some gravity experiments for the motion of binary neutron stars in strong external gravitational field. In the case of small velocity of the center of mass of the binary and the weak external gravitational field investigation of such systems can be performed by using the well known pulsar timing techniques (see, e. g., \cite{Teukolsky, Taylor, Wex, Tempo2, Damour-Deruelle}). The approaches to calculation of the times of arrival of pulses from pulsars that are moving in external gravitational field are discussed in some papers (see, e. g., \cite{Wang2008, Wang2009, Stovall2012, Angelil, Liu-Wex, Zhang2015}). But must of this are uses the post-Newtonian expansion of the times of arrival of pulses, that have large deviations from exact result for the motion of the source near the horizon of the supermassive black hole.
To study the motion in such strong field regimes it is necessary to improve of existing methods or develop of new approaches to this problem.

Another useful quantity is gravitational waves from the pulsar. The first direct detection of gravitational waves \cite{LIGO}  has shown to us an importance of the investigation of this characteristic of the binary \cite{LIGOGR}. Therefore, consideration of the problem of motion of binary systems
in the field of the supermassive Black Hole is very important for calculation of gravitational
and electromagnetic radiation from such binary systems.
Unfortunately, there is no
hope in any foreseeable future to have exact solutions describing the motion of three
massive bodies, so we have to adopt some sort of approximation schemes for solving the Einstein
equations in order to study such problems.

The equations of motion of isolated binary systems (see for example review \cite{Futamase}) are commonly derived by the means of Post-Newtonian expansion of Einstein equations in the powers of $v/c$ (Post-Newtonian approach), where $v$ is characteristic velocity of the bodies, $c$ is the vacuum speed of light, or in the powers of $G$ (Post-Minkowskian approach), where $G$ is the gravitational constant. Despite the fact, that 2-body problem has received considerable attention in the literature and has been solved up to 3.5PN order, the n-body problem is still much less investigated. It has been solved by Kopeikin \cite{Kopeikin} up to 1PN order under the assumption, that the fields are weak and the motion of bodies is non-relativistic.

It is clear that the problem of binary motion in the field of supermassive black hole may be solved by an approximate consideration of 3-body problem. Namely, the problem corresponds to the case $m_{1,2}/M\ll 1$, where $m_1$, $m_2$ are the masses of stars in the binary system and $M$ is the mass of the black hole. However, this approach seems to be inadequately complicated and in the case of relativistic motion of binary's center of inertia --- even more tedious. At the same time high mass of the black hole suggests rather simple approximate method. It is based on the fact that in the vicinity of the binary system there may be introduced a comoving reference frame, in which the equations of relative motion of the stars are close to Newtonian. The conditions under which the approximation is adequate will be formulated in Sec. \ref{Equation}, as well as numerical estimates for real binary systems of neutron stars.

In this paper we derive the equations of motion of the binary system that moves in external gravitational field. This equations can be applied to the any metric witch changes on a scale that is more large than the spatial size of the binary. Also we derive expression for the times of arrival of pulses that comes from pulsar in a binary system in external gravitational field. By using this expression it is possible to fit pulsar timing data to find parameters of motion.

The value of the angular momentum of the Galactic center black hole is not known quite exactly. This quantity most lies in the ranges of $0<a/M<1.$ But the formulas for the case of $a=0$ (Schwarzschild black hole) the formulas are much simpler than in general ($a\neq 0$). Thus give us possibilities to simply analyse the results that are needed for the solving of inverse problem (the obtaining of the parameters of motion by using the redshift data). Because of this we have analyzed this approach on the example of a source in a binary pulsar that moves near Schwarzschild black hole. The result can be applied to the analyzing timing data of the pulsar that moves in the vicinity of Sgr A*.

\section{Equations of motion  of a compact binary system in the field of the supermassive black hole   }\label{Equations_of_motion}
\subsection{Equations of motion in  a  comoving reference frame}\label{Equation}
It is known that in the general relativity the equations of motion of
a many-body system can be obtained from the Einstein field
equations. For the first time this idea had been realized by
Einstein and Grommer \cite{Einstein-Grommer}. It had received
further development by Einstein, Infeld, and Hoffmann \cite{Einstein},
Fock \cite{Fock}, Infeld and Pleba\'{n}ski \cite{Infeld},
Will \cite{Will} and many other authors. Using the method of
Einstein--Infeld--Hoffmann, we will derive the equations of motion of
the binary system in the field of the supermassive Black Hole. We
assume that the relative motion of the stars in this binary
system is non-relativistic (the motion of the binary system as a whole
relatively to the SBH can be relativistic or even ultrarelativistic). We can
simplify our calculations essentially by the use of the comoving
reference frame, i.e. the reference frame, which is connected to the
center of mass of the binary system.

Let us consider a gravitationally bound compact system which are freely moving in the field of a supermassive black hole.

Let us make the following assumptions about this system:
\begin{enumerate}
\begin{subequations}\label{S1}
\item
The mass $M$ of  the supermassive BH is much greater then the masses of the both stars
\begin{equation}\label{S1d}
    M\gg m_{1,2}\,.
\end{equation}
\item The mean distance $\varrho$  between the stars is much greater than their own sizes $ R_0$:
  \begin{equation}\label{S1a}
  \varrho\gg R_0
\end{equation}
(e.g.  for neutron stars  $\varrho\sim 4\pot{4}\kk{km}\,,\  R_0\sim 10\div20\kk{km}$).
It means that we can consider the stars in a good approximation  as point-like masses $m_1$ and $m_2$.
\item The relative motion of the stars with respect to each other is non-relativistic:
\begin{equation}\label{S1e}
    \frac{v}{c}\ll 1\,.
\end{equation}
\item The characteristic length scale of external field inhomogeneity is larger than the size of our binary star system.
    \begin{equation}\label{S1c}
        r_g\gg\varrho\,,
    \end{equation}
    where  $r_g=2M$ is the gravitation radius of the black hole.
\end{subequations}
\end{enumerate}

Under the assumptions \eqref{S1} gravitational radiation almost doesn't affect the orbital motion of binary (around black hole) as well as relative motion of the stars. Particularly, the estimates based on quadrupole formula show that relative decrease of the radius of circular orbit of binary neutron star in flat background due to gravitational radiation would be of order $\Delta \varrho/\varrho\sim 10^{-13} $ per period, if
\begin{equation}\label{S1b}
    \frac{v}{c}\sim 10^{-2}\quad\text{for}\quad
    \varrho\sim 4\pot{4}\kk{km}\,,\
    m_{1,2}\sim m_{\odot}\,,
\end{equation}
where $m_{\odot}$ is the mass of the Sun.
Hence the effects of gravitational radiation will not be taken into account in this paper. However, one must be aware that as the distance between the stars decreases to the order of 100 km, gravitational radiation causes rapid collapse of both stars onto each other \cite{Kuznetsov}. The assumptions \eqref{S1} allow us to simplify the calculations greatly by the use of a comoving reference frame.
\subsection{Comoving reference frame}
As a comoving reference frame  we  choose the reference frame of a
single observer \cite{Mitskievich}. This reference frame is determined by the motion of
a single mass point. The world line of this mass point
\begin{equation}\label{A1}
    x^{i}=\xi^i(\tau)\,,\quad \text{($\tau$  proper time)}
\end{equation}
(``single observer'') is named basis. Using the world line $\xi^i(\tau)$ of
the center of mass of the binary star as basis, we obtain a
convenient comoving reference frame for the binary star system. Let
us give a brief description of this  reference frame.

\begin{figure}
  \includegraphics[width=1\columnwidth]{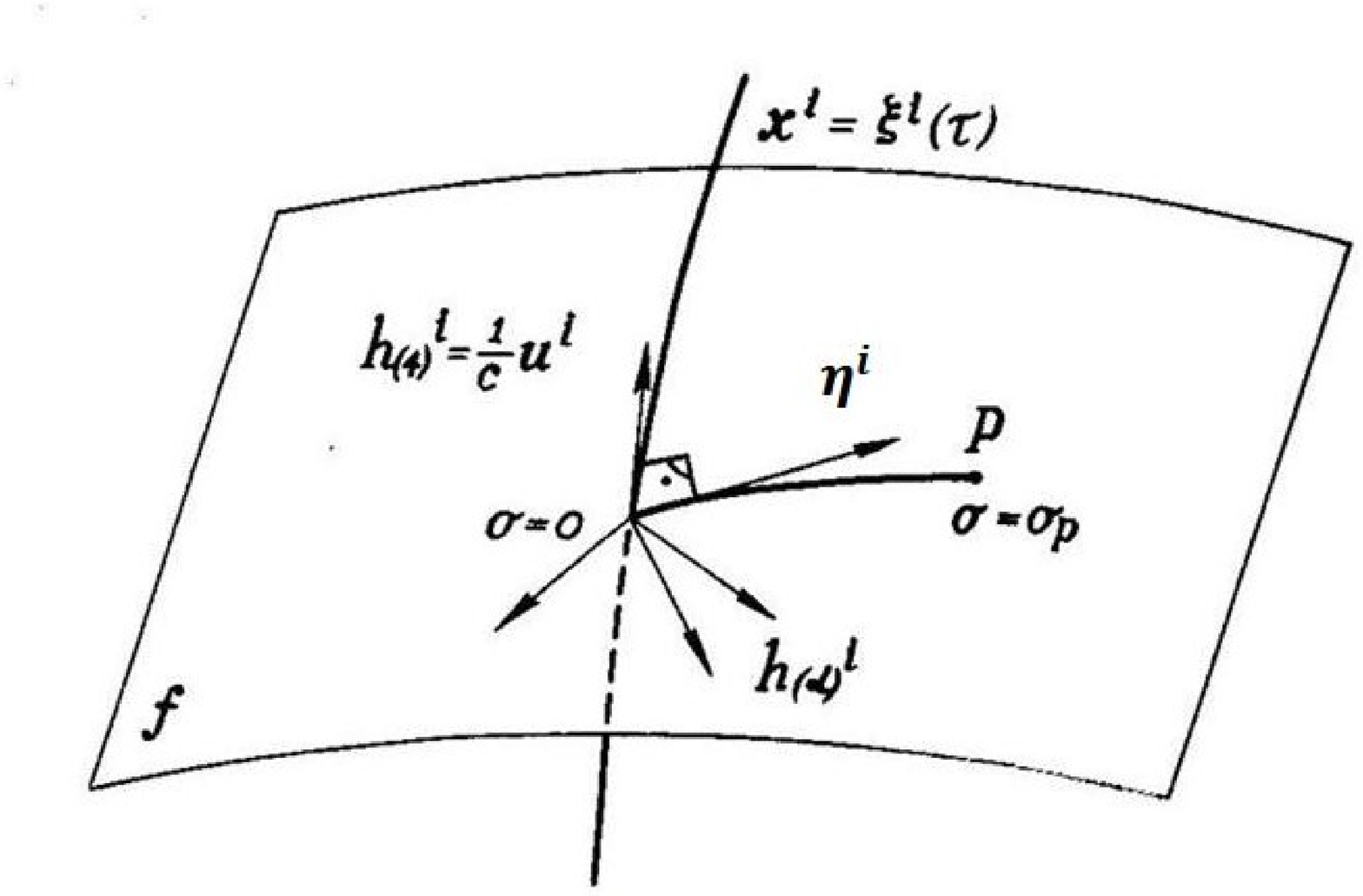}\\
  \caption{To the definition of generalized  Fermi coordinates}\label{fig2}
\end{figure}

Along the basic line $\xi(\tau)$ we establish an orthonormal
vierbein (tetrad) $h_{(m)}\w ^i$, defined by
\begin{equation}\label{A2}
    h_{(4)}\w ^i=\frac{1}{c}\,u^i\,,\quad u^i\equiv\deriv{\xi^i(\tau)}{\tau}\,,\quad
h_{(i)}\w ^k h_{(j)k}=\eta_{(i)(j)}\,,
\end{equation}
with $\eta_{(i)(j)} = \mathrm{diag}(1, 1, 1, -1)$ being the
Minkowski tensor and $c$ --- the speed of light\footnote{Latin indices run from 1 to
4, Greek ones from 1 to 3. The signature of space-time is $(+\,,
+,+,-)$. }.  The introduced
vierbein is determined up to three-dimensional
rotations. The three-dimensional physical space is given by a
geodesic spacelike hypersurface $f$ (related to $\tau$), which lies
orthogonally to the basic world line. In order to arithmetize the
hypersurface $f$, at each point $P\in f$ we fix a set of three
scalars
\[X^{(\alpha)}=\sigma_Ph^{(\alpha)}\w _i\eta^i\,,\]
where $\sigma_P$ is the
value of the canonic parameter $\sigma$ at $P$, defined along a
spacelike geodesic in $f$ and going through the point $P$, $\eta^i$ is
the tangent unit vector to that geodesic ($\eta_i\eta^i=1$), defined at
the point on the basis line ($\sigma=0$) (see Fig.~\ref{fig2}).
For a nonrotating frame
(i.e. when the vectors $h_{(n)}\w ^i$ are displaced along the basis line
\eqref{A1} according to the Fermi-Walker transport) the quantities
$\{X^{(\alpha)},c\tau\}$ correspond to the Fermi normal
coordinates (see for instance \cite{Mitskievich, Fortini}).
Analogous quantities
\begin{equation}\label{A4}
    x^{\hat{\alpha}}=X^{(\alpha)}\,,\;
x^{\hat{4}}=c\tau
\end{equation}
we treat as rotating Fermi coordinates. In these coordinates the
metric tensor $g_{\hat{i}\hat{j}}$ becomes
\begin{equation}\label{A5}
g_{\hat{i}\hat{j}}=\eta_{(i)(j)}+\varepsilon_{(i)(j)}\,,
\end{equation}
where $\eta_{(i)(j)}=\mathrm{diag}\,(1,1,1,-1)$ and
\begin{subequations}\label{A6}
 \begin{eqnarray}\label{A6a}
    &&\varepsilon_{(\alpha)(\beta)}=-\dfrac{1}{3}\,R_{(\alpha)(\mu)(\beta)(\nu)}
    X^{(\mu)}X^{(\nu)}+O(\varrho^3)\,,\\ \label{A6b}
    &&\varepsilon_{\alpha (4)}=\dfrac{1}{c}\,K_{\alpha}+\Theta_{\alpha}+O(\varrho^3)\,,\\ \label{A6c}
    &&\varepsilon_{(4)(4)}=-\Bigl(2\Theta+2\zeta+\zeta^2-\dfrac{1}{c^2}\,K_{\alpha}K^{\alpha}\Bigr)+O(\varrho^3)\,;\nonumber\\
    &&\varrho\equiv\sqrt{X^{\fa}X_{\fa}}\,.\nonumber
\end{eqnarray}
\end{subequations}
 \begin{eqnarray}
   &&K_{\fa}=\epsilon_{\fa\fk\ft}X^{\ft}\omega^{\fk}\,,\nonumber\\
    &&\zeta=\frac{1}{c^2}\,W_{\fa}X^{\fa}\,,\nonumber\\
    &&\Theta_{\fa}=\tfrac{2}{3}\,R_{\fa\fm\fn\fv}X^{\fm}X^{\fn}+O(\varrho^3)\,,\nonumber\\
&&\Theta=
\tfrac{1}{2}\,R_{\fv\fm\fv\fn}
X^{\fm}X^{\fn} +O(\varrho^3)\,.\nonumber
 \end{eqnarray}
Here, we used the following notations:
\begin{equation}\label{A8}
W^{\fa}=h^{\fa}\w _i\Deriv{u^i}{\tau}
\end{equation} and
\begin{equation}\label{A9}
\omega^{\fa}=\frac{1}{2}\,\epsilon^{\fa\fk\ft}h_{\ft
i}\Deriv{h_{\fk}\w ^i}{\tau}
\end{equation}
are the acceleration and the angular velocity of the reference
frame, respectively.

From the last relations it follows, that the size of a world tube in which
geodesic hypersurfaces are regular and in which the expansion \eqref{A5}
is valid, are determined from the following
conditions.
\begin{eqnarray}\label{A13}
    \varrho\ll\mathrm{Min}\Biggl\{&\dfrac{c^2}{\bigl|W_{(\nu)}\bigr|}\,,\;
\dfrac{c^2}{\bigl|\omega^{(\nu)}\bigr|}\,,\;
\dfrac{1}{\bigl|R_{(m)(n)(i)(j)}\bigr|^{1/2}}\,,\;\nonumber\\
&\hspace*{-2em}\dfrac{1}{\bigl|R_{(m)(n)(i)(j);(k)}\bigr|^{1/3}}\,,\;
\dfrac{\bigl|R_{(m)(n)(i)(j);(k)}\bigr|}{\bigl|R_{(m)(n)(i)(j);(k);(l)}\bigr|}\,\Biggr\}\,.
\end{eqnarray}
The present
calculation is carried out up to $\varrho^2$.

\subsection{Newtonian-like (non-relativistic) approximation in the comoving reference frame}
As the first step we shall consider the Newtonian-like
approximation in the comoving reference frame of a single
observer which had been described above. In particular,  in the Fermi coordinates $\{x^{\hat{i}}\}$
the metric tensor, describing a
gravitational field of SBH + binary star, will be sought in the
form
    \[g_{\hat{i}\hat{j}}=\eta_{(i)(j)}+\varepsilon_{(i)(j)}+\varphi_{(i)(j)}\]
where $\varepsilon_{(i)(j)} $ (background metric) is given by \eqref{A6}, and unknown
functions $\varphi_{(i)(j)}$ can be determined from Einstein
equations
\begin{equation}\label{A16}
    R_{\hat{i}\hat{j}}-\tfrac{1}{2}\,R
    g_{\hat{i}\hat{j}}=\kappa\,T_{\hat{i}\hat{j}}
\end{equation}
with usual expression for the stress-energy tensor
$T_{\hat{i}\hat{j}}$ describing two mass points $m_1$ and $m_2$.
Further we shall restrict our  consideration to non-relativistic
motion of this mass points (stars) relatively to each other. In this case
from equation \eqref{A16} we obtain immediately the Poisson-like
equation for non-relativistic relative motion of the stars
\begin{eqnarray*}
    &&\triangle \varphi(X^{\alpha})=4\pi G\Bigl(m_1\delta^{(3)}(X^{\alpha}-X_1^{\alpha})\\
    &&\!\!+ m_2\delta^{(3)}(X^{\alpha}-X_2^{\alpha})\Bigr)+
O(\varepsilon_{(i)(j)},\; 1/c^2)\,.
\end{eqnarray*}
Here $\varphi\equiv -c^2/2\,\varphi_{(4)(4)}$ is analogue of Newtonian potential.
The solution of the last equation corresponding to  boundary conditions has the shape
\begin{eqnarray*}
    \varphi(X^{\fa})=&&G\left(\frac{m_1}{\left|X^{\fa}-X^{\fa}_1\right|}+\frac{m_2}{\left|X^{\fa}-X^{\fa}_2\right|}\right)
    \\
    &&+O(\varepsilon_{(i)(j)}\,,\;1/c^2)\,.
\end{eqnarray*}

One can obtain the equations of motion of both stars from  the equation
\begin{equation}\label{A16a}
    T^{\hat{i}\hat{j}}{\!}_{; \hat{j}}=0\,,
\end{equation}
which follows from Einstein equations \eqref{A16}. Using the expression
\eqref{A16a} it is easy to show that this equations of motion can be
written as Lagrange equation with the following Lagrangian:
\begin{eqnarray}\label{A19}
    &&\mathcal{L}=\dfrac{m_1\Bm{v}_1^2}{2}+\dfrac{m_2\Bm{v}_2^2}{2}+\dfrac{Gm_1m_2}{|\Bm{r}_1-\Bm{r}_2|}\nonumber\\
    &&+
    \varepsilon_{\fa \fk
    \ft}\,\omega^{\fk}\left(m_1X_1^{\ft}v_1^{\fa}+m_2X_2^{\ft}v_2^{\fa}\right)\nonumber\\
    &&+\dfrac{2c}{3}\,R_{\fa \fm
    \fn(4)}\left[m_1X_1^{\fm}X_1^{\fn}v_{1}^{\fa}+m_2X_2^{\fm}X_2^{\fn}v_{2}^{\fa}\right]\nonumber
    \\
    &&
    -W_{\fa}\left(m_1X_1^{\fa}+m_2X_2^{\fa}\right)\notag\nonumber\\
    &&+D_{\fm\fn}\left(m_1X_1^{\fm}X_1^{\fn}+m_2X_2^{\fm}X_2^{\fn}\right)\,,
\end{eqnarray}
where the following abbreviations were used:
\begin{eqnarray*}
    D_{\fm\fn}&&=-\dfrac{c^2}{2}\,R_{(4)\fm(4)\fn}
   \\
    &&+\dfrac{1}{2}\,\left(\delta_{\fm\fn}\Bm{\omega}^2-\omega_{\fm}\omega_{\fn}
    \right),
\end{eqnarray*}
with $ \Bm{\omega}^2=\omega_{\fn}\omega^{\fn}$;
    \[v^{\fa}_{1,2}=\deriv{X^{\fa}_{1,2}}{T}\,.\]
Here  $T=1/c\,x^{\hat{4}}$ denotes  the time coordinate in the
comoving reference frame, i.~e. the proper time of observer
\eqref{A4}, which coincides with the proper time of the
center of mass of the binary system.

After the transformation
\begin{eqnarray*}
\left\{
\begin{aligned}
  &X_{1}^{\fa}=X^{\fa}+\dfrac{m_2}{m_1+m_2}\,x^{\fa}\\
  &X_{2}^{\fa}=X^{\fa}-\dfrac{m_1}{m_1+m_2}\,x^{\fa}
\end{aligned}\right.
\end{eqnarray*}
and
\begin{eqnarray*}
\left\{
\begin{aligned}
  &v_{1}^{\fa}=V^{\fa}+\dfrac{m_2}{m_1+m_2}\,v^{\fa}\\
  &v_{2}^{\fa}=V^{\fa}-\dfrac{m_1}{m_1+m_2}\,v^{\fa}
\end{aligned}\right.\,.
\end{eqnarray*}
into the reference frame of Newtonian  center of mass (in the Fermi coordinates) we obtain for the Lagrangian \eqref{A19}
\begin{eqnarray}\label{A21}
    &&\mathcal{L}=\Bigl[\dfrac{\m \Bm{v}^2}{2}+\dfrac{G
    m_1m_2}{r}+\m\Bigl(\varepsilon_{\fa\fk\ft}\omega^{\fk}
    x^{\ft}v^{\fa}\notag\nonumber\\
    &&+\dfrac{2c}{3}\,R_{\fa \fm
    \fn(4)}\dfrac{m_2-m_1}{m_1+m_2}\,
    x^{\fm}x^{\fn}v^{\fa}\notag\nonumber\\
    &&+D_{\fm\fn}x^{\fm}x^{\fn}\Bigr)\Bigr]\\
    &&+
    (m_1+m_2)\Bigl[\dfrac{1}{2}\,\Bm{V}^2+\varepsilon_{\fa \fk
    \ft}\omega^{\fk}X^{\ft}V^{\fa}\notag\nonumber\\
    &&+\dfrac{2c}{3}\,R_{\fa \fm \fn
    (4)}X^{\fm}X^{\fn}V^{\fa}+D_{\fm\fn}X^{\fm}X^{\fn}\notag\nonumber\\
    &&-W_{\fa}X^{\fa}\Bigr]+
    \dfrac{\m c}{3}\,\Bigl(R_{\fa\fm
    \fn(4)}+R_{\fa\fn\fm(4)}\Bigr)\notag\nonumber\\
    &&\quad\times\Bigl(2x^{\fn}X^{\fm}v^{\fa}+x^{\fm}x^{\fn}v^{\fa}\Bigr)\,,\nonumber
\end{eqnarray}
where
\begin{eqnarray}\label{A22}
{\Bm{\varrho}}&&=\Bm{r}_1-\Bm{r}_2\,, \quad
x^{\fa}=X_{1}^{\fa}-X_{2}^{\fa}\,,\quad\text{and}\nonumber\\
\varrho &&=|\Bm{\varrho}|=\sqrt{x^{\fa}x_{\fa}}\,,
\end{eqnarray}
    \begin{eqnarray*}
    &&\Bm{R}=\dfrac{m_1
\Bm{r}_1+m_2 \Bm{r}_2}{m_1+m_2}\quad\text{and}\\&&
X^{\fa}=\dfrac{m_1 X_1^{\fa}+m_2 X_{2}^{\fa}}{m_1+m_2}\,,
\end{eqnarray*}
    \[\Bm{v}=\deriv{\Bm{\varrho}}{T}\,,\quad\text{and}\quad
    \Bm{V}=\deriv{\Bm{R}}{T}\,,\]
    \[\m=\frac{m_1 m_2}{m_1+m_2}\,.\]
The equations of motion of the center of mass (equation for $\Bm{R}$)
and the equations of motion of both stars (equation for
$\Bm{\varrho}$) relative to each other can be written in the
Lagrange form
\begin{subequations}\label{A26}
    \begin{eqnarray}\label{A26a}
&\deriv{}{T}\left(\pderiv{\mathcal{L}}{V^{\fa}}\right)-\pderiv{\mathcal{L}}{X^{\fa}}=0
\quad\text{and}
\\ \label{A26b}
&\deriv{}{T}\left(\pderiv{\mathcal{L}}{v^{\fa}}\right)-\pderiv{\mathcal{L}}{x^{\fa}}=0\,,
\end{eqnarray}
\end{subequations}
respectively. After simple calculations we obtain from \eqref{A26}
\begin{eqnarray}\label{A26A1}
    (m_1&&+m_2)\,\deriv{V^{\fk}}{T}=
    (m_1+m_2)\bigl(2\varepsilon^{\fk}\w
    _{\fm\fn}V^{\fm}\omega^{\fn}\nonumber\\
    &&-2cR^{\fk}\w _{\fn\fm
    (4)}X^{\fm}V^{\fn}+2 D^{\fk}\w _{\fn}
    X^{\fn}-W^{\fk}\bigr)\nonumber\\
    && -2m^*cR^{\fk}\w _{\fn\fm (4)}x^{\fm} v^{\fn}\,,
\end{eqnarray}
\begin{eqnarray}\label{A26A2}
    &&\deriv{v^{\fk}}{T}=
    \Bigl(\frac{G(m_1+m_2)}{r}\Bigr)_{\Sp \fk}-2    \varepsilon^{\fk}\w _{\fa\ft}\omega^{\fa}
    v^{\ft}\nonumber\\
    &&\qquad-\frac{2c(m_2-m_1)}{(m_1+m_2)}\,R^{\fk}\w _{\fn\fm (4)}x^{\fm}
    v^{\fn}
     +2 D^{\fk}\w _{\fm} x^{\fm}\nonumber\\
    &&\qquad +\frac{4}{3}\,c
    S_{\fa\fm\fk}\bigl(X^{\fm} v^{\fa}+x^{\fm}
    V^{\fa}\bigr)\,,
\end{eqnarray}
where the abbreviation
\begin{equation}\label{A26A3}
    S_{\fa\fm\fn}=\frac{1}{2}\,\bigl(R_{\fa\fm\fn
    4}+R_{\fa\fn\fm
    (4)}\bigr)
\end{equation}
is used.

In order to use the reference frame, comoving with the center of mass,
 we let
\begin{equation}\label{A27}
    X^{\fa}=0\,,\quad V^{\fa}=0\,.
\end{equation}
It is obvious that these conditions will make sense, if the equation $\mathrm{d}V^{\fa}/\mathrm{d}t=0$
will follow from \eqref{A27} and  \eqref{A26a}. Taking into account the equation \eqref{A26A1} (i.e.
the explicit form of equation \eqref{A26a}), we obtain the
following expression for 4-acceleration of the center of masses
in the comoving Fermi coordinates:
\begin{eqnarray*}
    W^{\fa}=&&-\frac{c}{m_1+m_2}\,R^{\fa}{\!}_{\fm\fn
    (4)}\varepsilon^{\fm\fn}\
    _{\fb}M^{\fb}\\
    &&-\frac{c}{3(m_1+m_2)}\,\deriv{}{t}\left(R^{\fa}\
    _{\fm\fn(4)}Q^{\fm\fn}\right)\,.
\end{eqnarray*}
In the above formula,
\begin{eqnarray*}
    M^{\fa}&&=m_1\varepsilon^{\fa}\w_{\fk\ft}X^{\fk}_1v^{\ft}_1+
    m_2\varepsilon^{\fa}\w{}_{{\fk\ft}}X^{\fk}_2v^{\ft}_2\\
    &&=\m \varepsilon^{\fa}\w_{\fk\ft}x^{\fk}v^{\ft}
\end{eqnarray*}
is the intrinsic angular momentum of the binary, which is
calculated with respect to its center of mass,
\begin{eqnarray*}
    Q^{\fm\fn}&&=\sum\limits_{a=1}^{2}m_{a}\left(3X_{a}^{\fm}X_{a}^{\fn}-\Bm{r}^2_{a}\delta^{\fm\fn}\right)
    \\ &&=
    \m \left(3x^{\fm}x^{\fn}-\Bm{\varrho}^2\delta^{\fm\fn}\right)\quad
    \text{(if $R^{\fa}=0$)}
\end{eqnarray*}
denotes the quadrupole moment tensor. Let's notice that in the approximation used
here we can put
    \[\deriv{}{T}\,\bigl(R^{\fa}\w _{\fm\fn(4)}Q^{\fm\fn}\bigr)=
    R^{\fa}\w _{\fm\fn(4)}\deriv{}{T}\,Q^{\fm\fn}\,,\]
\begin{eqnarray*}
&\deriv{}{T}\,\bigl(S_{\fa\fm\fn}x^{\fn}X^{\fm}x^{\fa}\bigr)\\& \qquad=
    S_{\fa\fm\fn}\deriv{}{T}\bigl(x^{\fn}X^{\fm}x^{\fa}\bigr)
\end{eqnarray*}
and so    on.
In other words one can say that the center of mass of the binary
star satisfies in good approximation the following equations \cite{Gor1,Gor2}
\begin{eqnarray}\label{A31}
    (m_1+m_2)\Deriv{u^i}{\tau}&&=-\frac{1}{2c}\,R^{i}\w{}_{skm}u^s
    \varepsilon^{mkbn}
    {M}_{b}u_n\nonumber
    \\
    &&-\frac{1}{3}\,h^i\w{}_{s}\Deriv{}{\tau}\left(R^s\w{}_{klm}Q^{kl}u^{m}\right)\,,
\end{eqnarray}
where $\varepsilon^{mkbn}$ is the Levi-Civita pseudotensor
($\varepsilon^{1234}=1/\sqrt{-g}$; $g=\det(g_{ij})$); $h^i\w{}_{s}=g^i\w{}_{s}+(1/c^2)u^iu_s$ is the projective tensor.
Taking into account the conditions \eqref{A27}, thus we obtain from
\eqref{A26A2} the following equation of relative motion in
Newtonian approximation and in the comoving Fermi coordinates \cite{Gor1}:
    \[\deriv{\Bm{v}}{T}=-\nabla(\phi+\Upsilon)+2\Bm{v}\times\Bm{\omega}+\Bm{A}\,,\]
where
\begin{eqnarray*}
    \Upsilon=&&-D_{\fa\fb}x^{\fa}x^{\fb}=\frac{c^2}{2}\,R_{(4)\fa(4)\fb}x^{\fa}x^{\fb}\\
    &&-\frac{1}{2}\,\Bigl[\Bm{\varrho}^2\Bm{\omega}^2-
    (\Bm{\varrho}\cdot\Bm{\omega})^2\Bigr]\,,
\end{eqnarray*}
    \[\quad\phi=-\frac{G(m_1+m_2)}{r}\,,\]
\begin{eqnarray*}
    A^{\fa}=&&\frac{c(m_1-m_2)}{m_1\,m_2}\,\Bigl[R^{\fa}\w _{\fm\fn
    (4)}\varepsilon^{\fm\fn}\w
    _{\fa}M^{\fa}\\
    &&-\frac{1}{3}\,\deriv{}{T}\left(R^{\fa}\w
    _{\fm\fn(4)}Q^{\fm\fn}\right)\Bigr]\,.
\end{eqnarray*}

It should be noticed that equations of motion \eqref{A31}, derived in this section,
are in accordance with those obtained by J.~Anandan by the use of action based approach~\cite{Anandan1,Anandan2}.

\section{Redshift\label{Sredshift}}
Electromagnetic radiation is a unique source of information about the motion of compact binaries in external gravitational field.

A typical wavelength of radiation used for observations $\lambda\lesssim 10^{3}\text{m}$ is much less than the scale of gravitational inhomogeneities $M\sim 10^{9}\text{m}$. Because of this we use the geometrical optics approximation (see e. g.  \cite{Stephani}). We will consider two characteristics of electromagnetic radiation:  times of arrival of the pulses $t_{TOA}$ and redshift $z$. Times of arrival is the moment of observation of pulses of pulsar and these is commonly used in the analysis of pulsar timing (see e. g. \cite{ObservationTimeD, PNtiming, Teukolsky, Wex, 21, Hulse, HulseT}). Radiation of usual stars (main sequence stars, white dwarfs or giants) is usually described by redshift (see e. g. \cite{Stars, HMCancri, Tarasenko, KerrRedShift}). Redshift is related to the times of arrival as follows (see \cite{Sing}):
\begin{equation}\label{observable}
\frac{\delta \lambda}{\lambda}=z=\frac{t_{TOA}-t_{TOA'}}{T_p}-1.
\end{equation}
where $\lambda$ is the wavelength of the emitted light, $\delta\lambda$ is the difference between wavelengths of the arrival light and the emitted light, $t_{TOA}-t_{TOA'}$ is the difference between times of arrival of two consistent pulses of the pulsar, $T_p$ is the period of the pulsar in the pulsar reference frame. Because the redshift $z$ and difference of the times of arrival $t_{TOA}-t_{TOA'}$ are interconnected, we can choose the former as radiation characteristic.

Redshift of a radiation source which moves in external gravitational field can be calculated using the following expression (see \cite{Sing}):
\begin{equation}\label{generalredshift}
z+1=\frac{(k^iu_i)_s}{(k^iu_i)_o},
\end{equation}
 where $u^i$ is the 4-velocity vector of the source (subscript "s") or the observer (subscript "o"), and $k^i$ is the wave vector of the ray in the corresponding points.

From observations we know redshift as a function of the observer time: $z=z_t(t).$ But it is more convenient in calculations to use the redshift as a function of proper time of the source $z(\tau)$. The transition from the function $z(\tau)$ to the $z_t(t)$ can be accomplished by using the following expression:
\begin{equation}
t(\tau)=\int\limits^{\tau}_0\frac{d\tau}{z(\tau)+1},
\end{equation}
where we chose the initial proper time of the source such that $t(0)=0.$ Usually the obtained function is monothonic and we can find inverse function $\tau(t).$  Then, we have
\begin{equation}
z_t(t)=z(\tau(t)).
\end{equation}
Therefore for calculation of the redshift it is enough to know only the function $z(\tau).$

Since the observer is far from the Galactic Center, we can use the Minkowski metric and the Galilean coordinates $(t, x^{\alpha})$ in the vicinity of observer. Then, we obtain
\begin{equation}\label{Earth}
\frac{(k^iu_i)_o}{A}\approx\frac{c+n_{\alpha}v^{\alpha}}{\sqrt{1-v_{\alpha}v^{\alpha}/c^2}},
\end{equation}
where $v^{\alpha}=dx^{\alpha}/dt,$ $n_{\alpha}$ is the unit vector in the direction of the Galactic Center. $A$ is integral of motion and it quantity is dependent only on the parametrization of the ray. Due to this without loss of generality we can establish $A=1$ (see Appendix A). Then the redshift can be expressed as
\begin{equation}\label{redshift}
z+1=\frac{z_{\infty}+1}{(k^iu_i)_o}.
\end{equation}
In the formula (\ref{redshift}) only the part $z_{\infty}+1$ is of our interest. If one know this part, the whole expression for the redshift (\ref{generalredshift}) can be obtained by using the ephemerides of the Earth. The $z_{\infty}$ can be expressed as
\begin{equation}\label{redshiftfinal}
z_{\infty}+1=(k^i u_i)_s.
\end{equation}

The influence of the external gravitational field on the binary can be approximately described by an effective potential $u_{eff}=m_{1}c^2\cdot R_{(\alpha)(4)(\beta)(4)}x^{(\alpha)}x^{(\beta)}$ (see formula (\ref{A26A2})). For the system to be stable, this potential must be much less than the newtonian potential. In the present work we assume that
\begin{equation}\label{potentials}
|m_{1}c^2\cdot R_{(\alpha)(4)(\beta)(4)}x^{(\alpha)}x^{(\beta)}|\ll G\dfrac{m_{1}m_{2}}{r}.
\end{equation}
Therefore
\begin{equation}\label{periods}
\frac{T}{T_0}\ll 1,
\end{equation}
where $T_0$ and $T$ are characteristic timescales of motion of the binary in external field and relative motion of the stars in binary, respectively.

It is follows from (\ref{periods}) that the whole redshift is consists of two parts: the first one is fast oscillate (on timescale $\sim T$) and with relatively small magnitude (of the order $v$), and the second one is changes on timescale of $T_0$ and has the order of magnitude of 1 (see Sec. \ref{direct}).

\section{Schwarzschild metric}
\label{SchwarzschildS}
Let us consider the case of external gravitational field of a Schwarzschild black hole to apply formalism that have been developed in this work. This field can be used as an approximation of the gravitational field of the supermassive black hole in the Galactic Center.

The Schwarzschild metric has the following form (see e. g. \cite{Ch}):
\begin{eqnarray}\label{Schwarzschild}
&&g_{ij}{}^{Sch} dx^i dx^j=\mathrm{d}s^2=\\
&&\dfrac{\mathrm{d}r^2}{1-2M/r}+r^2 \mathrm{d}\theta^2+r^2\sin^{2}\theta \mathrm{d}\phi^2-\left(1-\dfrac{2M}{r}\right) \mathrm{d}(ct)^2,\nonumber
\end{eqnarray}
where $(r,\theta,\phi,t)$ are the Schwarzschild coordinates.

Components of 4-velocity of a timelike geodesic can be written as \cite{Ch}: 
\begin{subequations}
\begin{eqnarray}
&u^{3}=\dfrac{\mathrm{d}\phi}{\mathrm{d}\tau}=\dfrac{Lc}{r^2},
\label{u^{3}}\\
&\dfrac{1}{c}u^{4}=\dfrac{\mathrm{d}t}{\mathrm{d}\tau}=\dfrac{E}{1-2M/r},
\label{u^{4}}\\
&u^{\theta}=0,\\
&\dfrac{1}{c}u^{1}=\pm\sqrt{E^2-\left(1-\dfrac{2M}{r}\right)\left(1+\dfrac{L^2}{r^2}\right)}.
\label{u^{1}}
\end{eqnarray}
\end{subequations}
where $E$ is mechanic energy and $L$ is the angular momentum per unit mass of the test particle. In the chosen coordinate system $\theta(\tau)=\pi/2$.

4-wave vector  $k^{i}$ of an isotropic geodesic is given by (we choose the parametrization with $A=1$, see Sec. \ref{Sredshift}):
\begin{subequations}
\begin{eqnarray}\label{k^{3}}
&k^{3}=\dfrac{D}{r^2}\dfrac{1}{c};\\
&k^{4}=\dfrac{1}{(1-2M/r)c};\label{k^{4}}\\
\label{k^{2}}
&k^{\theta}=0;\\
&k^{1}=\pm\sqrt{\left[1-\dfrac{D^2}{r^2}\left(1-\dfrac{2M}{r}\right)\right]\dfrac{1}{c^2}}.\label{k^{1}}
\end{eqnarray}
\end{subequations}
where integral of motion $D$ is the impact parameter of the ray. Taking into account (\ref{k^{3}}), (\ref{u^{3}}) and  (\ref{k^{1}}), from (\ref{u^{1}}) we get
\begin{itemize}
\item The timelike geodesic
   \begin{eqnarray}\label{timeg}
   &&\dfrac{1}{r_s(\phi_s)}=\dfrac{1-e}{\ell}+\dfrac{2e}{\ell}\sn^{2}\left(\dfrac{(\phi_{s}-\phi_{s0})\sqrt{1-6M/\ell+2eM/\ell}}{2},k\right);\nonumber\\
   &&k=\sqrt{\frac{4\cdot e}{\ell/M-6+2e}}.
   \end{eqnarray}
\item The isotropic geodesic
  \begin{eqnarray}\label{lightg}
   &&\dfrac{1}{r_r(\phi_{r})}=\dfrac{1}{P}-\dfrac{Qk^2}{2PM}\nonumber\\
   &&\times\cn^{2}\left(\left[\dfrac{\phi_{r}}{2}\sqrt{\dfrac{Q}{P}}+\F\left(\arccos\left(\sqrt{\dfrac{2M}{Qk^2}}\right),k\right)\right],k\right);\nonumber\\
   &&k=\sqrt{\dfrac{Q-P+6M}{2Q}}.
   \end{eqnarray}
\end{itemize}
where $\F[\phi,k]$ is the elliptic integral of the first kind. Here the lower indices $r$ and $s$ denote the light ray and the radiation source, respectively. Angles $\phi_{r}$ and $\phi_{s}$ are measured in the planes of the ray and the source, respectively, as shown in Fig. \ref{Orbits2e}. $\phi_{s0}$ --- some initial angle of the orbit. The initial point for the isotropic geodesic is chosen to be at spatial infinity where $\phi_r=0$.
Formula (\ref{lightg}) is valid only for orbits that have pericenter, which corresponds to $D>3\sqrt{3}M$. The integrals of motion are related to the apocenter distance $s_{1}$  and pericenter distance $s_{2}$:
\begin{eqnarray}
   \left\{
     \begin{aligned}
    &L=\frac{s_1 s_2\sqrt{2}}{\sqrt{(s_2/M-2)(s_1^2+s_1 s_2)-2s_2^2}};\nonumber\\
    &E=\sqrt{\frac{(s_2-2M)(s_1-2M)(s_1+s_2)}{(s_2-2M)(s_1^2+s_1 s_2)-2Ms_2^2}}.
     \end{aligned}\label{EL}
    \right.
\end{eqnarray}
Also we use the abbreviations:
\begin{eqnarray}
      &\ell=\dfrac{2 s_{1} s_{2}}{s_{1}+s_{2}};\nonumber\\
      &e=\dfrac{\ell}{2}\dfrac{s_{1}-s_{2}}{s_{1} s_{2}};\\
      &Q=\sqrt{P^{2}+4MP-12M^2}.\nonumber
  \end{eqnarray}

\section{Calculation of the redshift}
\label{direct}
The purpose of this section is to describe a method of calculation of the redshift of the source in binary star system that moves in gravitational field of supermassive black hole.
Redshift $z$ of a point source of radiation is given by (\ref{redshiftfinal}). To apply this formula it is necessary to know the low of motion of the source. It follows as the solution of equations (\ref{A26A1}) and (\ref{A26A2}).

As a first approximation it is possible to solve numerically the equations (\ref{A26A2}) in assumption $X^{(\alpha)}=0,$ $V^{(\alpha)}=0.$  For the following approximations it is not difficult to solve the equations of motion of the center of mass (\ref{A26A1}). But from it is follows from the analysing of equations (\ref{A26A1}) that the right-hand side of it is negligible (has the order of $\rho v/r^2$) and we can consider the motion of the center of mass as geodesic ($X^{(\alpha)}=0,$ $V^{(\alpha)}=0.$)

Let us denote $z_0$ the redshift on infinity of a (non-real) source moving along the world line of the center of mass of the binary system. The redshift of light from this source can be found as
\begin{equation}\label{z0}
1+z_0=k_iu^i.
\end{equation}
Where the components of the velocity vector are calculated from (\ref{u^{1}}), (\ref{u^{3}}), (\ref{u^{4}}). The binary star system must be described as the finite-size object. Therefore the redshift for the source in binary system is (see Appendix B)
\begin{equation}\label{zz}
1+z_{\infty}=(1+z_0)\left(1-\frac{d}{d\tau}\left(n_{(\alpha)}X_1^{(\alpha)}\right)\right)+O(\rho^2).
\end{equation}
Here $X_1^{\alpha}(\tau)=-x^{\alpha}(\tau)m_2/(m_1+m_2)$ (we consider that the radiation of the component with index 1 is observed) are the vierbein components of the deviation of the source trajectory from the world line $\xi(\tau).$ The coordinate functions  $x^{\alpha}(\tau)$  must be calculated from (\ref{A26A2}).

\begin{figure}
\includegraphics[width=0.7\linewidth]{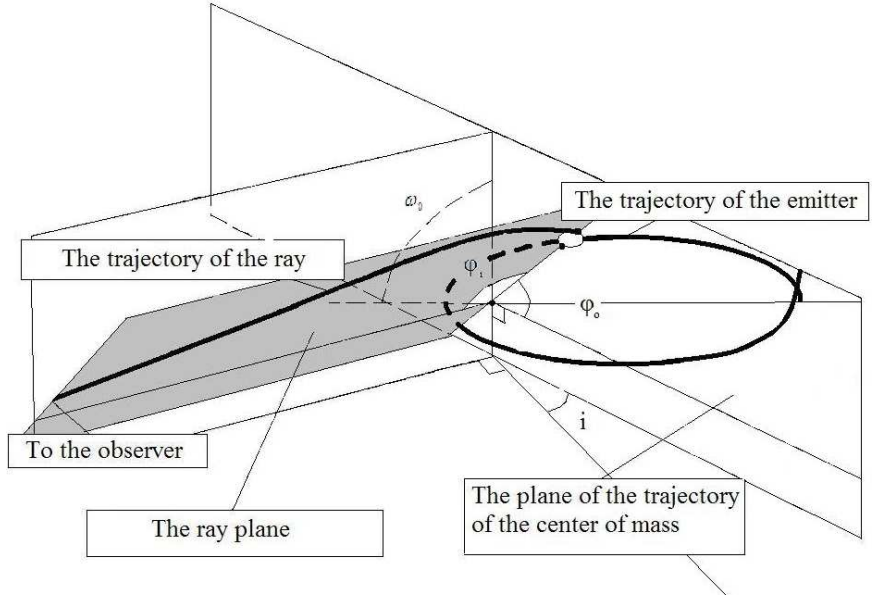}
\caption{Orbital parameters of the binary:}
$\phi_{0}$ ---  the polar angle in the orbital plane;
$\phi_{r}$ --- the polar angle in the ray plane;
$i$ --- the orbital inclination;
$\omega_{0}$ --- the longitude of periastron.\label{Orbits2e}
\end{figure}

To describe the orientation of the orbit of the center of mass, we use the orbital inclination $i$ and the longitude of periastron $\omega_0$ (see Fig. \ref{Orbits2e}). These two angles together with the integrals of motion $s_1$, $s_2$ form a full set of parameters of motion of the center of mass.

We have the following expression for the angle $\phi_r$:
\begin{equation}\label{phi_r}
\phi_r=\arccos(\cos\phi_s\sin i).
\end{equation}

It is convenient to choose the following vierbeins:
\begin{subequations}\label{ts}
\begin{eqnarray}
   \left\{
     \begin{aligned}
     &h_{(4)}{}^{1}=\left(\sqrt{E^2-\left(1-\dfrac{2M}{r}-\dfrac{2ML^2}{r^3}\right)-\dfrac{L^2}{r^2}}\right)\\
     &\times\left(\sign\left(\dfrac{dr}{d\tau}\right)\right); \\  &h_{(4)}{}^{2}=0;\\  &h_{(4)}{}^{3}=\dfrac{L}{r^2}; \\  &h_{(4)}{}^{4}=\dfrac{Er}{(r-2M)}.\label{4t}
     \end{aligned}
    \right.
\end{eqnarray}
\begin{eqnarray}
   \left\{
     \begin{aligned}
       &h_{(1)}{}^{1}=L\sqrt{\dfrac{E^2r^3-(r-2M)(r^2+L^2)}{L^2r^3+r^5}}\\
       &\times\left(\sign\left(\dfrac{dr}{d\tau}\right)\right); \\ &h_{(1)}{}^{2}=0;\\&h_{(1)}{}^{3}=\dfrac{\sqrt{L^2+r^2}}{r^2};  \\&h_{(1)}{}^{4}=\dfrac{LEr}{(r-2M)\sqrt{L^2+r^2}}.\label{1t}
     \end{aligned}
    \right.
\end{eqnarray}
\begin{eqnarray}
   \left\{
     \begin{aligned}
      &h_{(2)}{}^{1}=-\dfrac{Er}{\sqrt{L^2+r^2}}; \\ &h_{(2)}{}^{2}=0; \\ &h_{(2)}{}^{3}=0; \\ &h_{(2)}{}^{4}=\left(-\dfrac{\sqrt{E^2r^4-r(r-2M)(r^2+L^2)}}{(r-2M)\sqrt{L^2+r^2}}\right)\\
      &\times\left(\sign\left(\dfrac{dr}{d\tau}\right)\right).
     \end{aligned}
    \right.
\label{2t}
\end{eqnarray}
\begin{equation}
h_{(3)}{}^{i}=\left\{0,-\frac{1}{r},0,0\right\}.
\label{3t}
\end{equation}
\end{subequations}
Angular velocity (\ref{A9}) has one non-zero component
\begin{equation}
\omega=\omega^{(3)}=\frac{LE}{L^2+r^2}.
\label{om}
\end{equation}

The non-zero components of curvature tensor are:
\begin{eqnarray}
      &&R_{(2)(4)(2)(4)}=-\dfrac{3L^2+2r^2}{r^5},\nonumber\\&&R_{(3)(4)(3)(4)}=\dfrac{3L^2+r^2}{r^5},\nonumber \\&&R_{(1)(4)(1)(4)}=\dfrac{1}{r^3}, \\&&R_{(1)(3)(3)(4)}=-\dfrac{3L\sqrt{L^2+r^2}}{r^5}=-R_{(1)(2)(2)(4)}.\nonumber
\end{eqnarray}

For a given trajectory of the center of mass one knows the integrals $L$ and $E$. To determine the impact parameter of the ray it is necessary to solve the boundary value problem. In our case it reduces to the following non-linear equation
\begin{equation}
r_r(\phi_r(\phi_s))=r_s(\phi_s).
\end{equation}
By using representations of radial functions (\ref{timeg}) and (\ref{lightg}), one can find impact parameter $D$ for all $\phi_s$.
The components of the 4-wave vector $k^i$ in a reference frame rotating relative to the Schwarzschild reference frame are:
\begin{eqnarray}\label{realk}
&k'^{1}=k^1;\nonumber\\
&k'^{2}=-k^3\sin\Omega;\nonumber\\
&k'^{3}=k^3\cos\Omega;\\
&k'^{4}=k^4.\nonumber
\end{eqnarray}
where the components $k^i$ are given by (\ref{k^{1}}), (\ref{k^{3}}), (\ref{k^{4}}), $\Omega$ is the angle between the ray plane and the orbital plane of the center of mass motion:
\begin{equation}\label{Omega}
\Omega=\arccos\left(\frac{\sin\phi_s\sin i}{\sin\phi_r}\right).
\end{equation}
The redshift from the motion of the center of mass $z_0$  has the form:
\begin{eqnarray}\label{RedShiftCM}
   &z_{0}=\pm\sqrt{\left(\dfrac{E^2r}{r-2M}-(1+L^2/r^2)\right)}\nonumber\\
   &\times\sqrt{\left(\dfrac{r}{r-2M}-P^3/(r^2(P-2M))\right)}\nonumber\\
   &+\dfrac{L\sqrt{P^3}}{r^{2}\sqrt{P-2M}}\cos{\Omega}+\dfrac{E}{1-2M/r}-1.\nonumber\\
\end{eqnarray}
The vierbein components of the wave vector has the form $k^{(\alpha)}=h^{(\alpha)}{}_i k'^{i}$.

We find the redshift as a function of $\phi_s$ and $\tau$. In order to calculate redshift as a function of proper time of the source one must solve a differential equation

\begin{equation}
\frac{d\phi_s}{d\tau}=\frac{L}{r_s^2(\phi_s)}.
\end{equation}
This equation for the function $\phi_s(\tau)$ can be solved numerically.

To find the relative motion of the stars, we have solved equations (\ref{A26A2}) numerically. Choosing parameters of relative motion as in Table \ref{parametersofmotion}, it is simply to show that conditions (\ref{S1e}) and (\ref{S1c}) are satisfied: after numerical calculations we have obtained $v/c<0,01,$ $\rho/M<0,02$.    An example of calculating of the redshift is presented on Figure \ref{redshift}. The parameters of motion have been used are summarised in Table \ref{parametersofmotion}.
\begin{figure}
\includegraphics[width=0.7\linewidth]{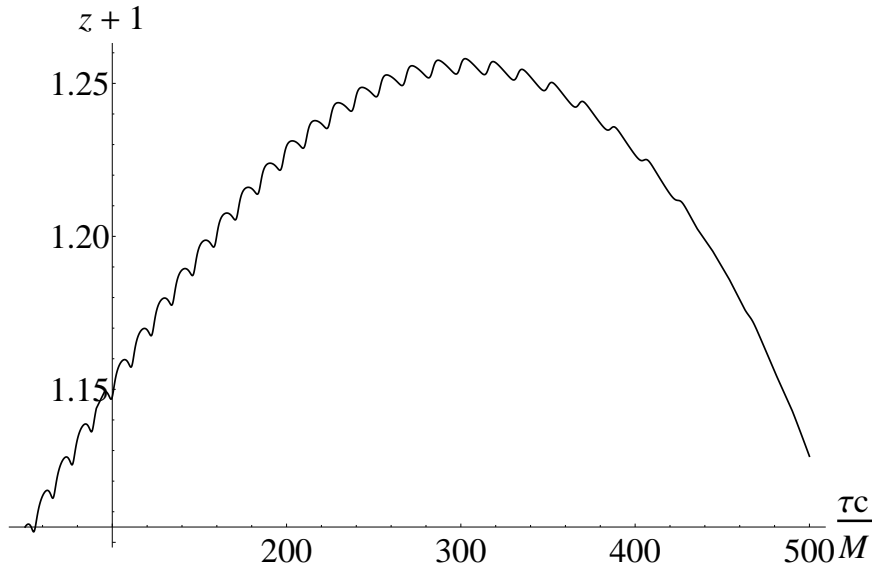}
\caption{The redshift as a function of the proper time of the source.}\label{redshift}
\end{figure}
\begin{table}
\begin{ruledtabular}
\caption{Parameters of the binary system, that is used to calculate the redshift (Figure \ref{redshift})}\label{parametersofmotion}
\begin{tabular}{l c}
   Parameter & Model value\\
\hline
 Pericenter distance, $s_1/M$ & 20,0 \\
Apocenter distance, $s_2/M$ & 30,0 \\
 Orbital inclination, $i$ &  1,00 \\
 Mass of the source, $m_1/M$ & $8,89\cdot 10^{-7}$\\
 Mass of the companion star, $m_2/M$ & $4,45\cdot 10^{-7}$\\
 Initial relative distance, $\left\{\dfrac{x^{(1)}}{M};\quad\dfrac{x^{(2)}}{M};\quad\dfrac{x^{(3)}}{M}\right\}$ &  $\{0;\quad 0,02;\quad 0\}$\\
 Initial relative velocities, $\left\{\dfrac{v^{(1)}}{cM};\quad\dfrac{v^{(2)}}{cM};\quad\dfrac{v^{(3)}}{cM}\right\}$ &  $\{0,0083;\quad 0;\quad 0,0057\}$\\
 \end{tabular}
\end{ruledtabular}
\end{table}

\section{The magnification factor and extinction of pulses of the pulsar}
Let us consider the times of arrival of pulses of a pulsar in binary system that moves in the gravitational field of a supermassive black hole. The times arrival of these pulses can be simply obtained from the redshift (\ref{observable}).
Due to the precision of the rotation axis of the pulsar and the deviation of the wave vector in the curved space-time an observer can see the pulses in a finite intervals of time (see e. g. \cite{Wang2009, Stovall2012}). The formalism that has been developed in this paper give us possibilities to calculate this time intervals on timescal of several periods of orbital motion. 

We introduce a spherical system of coordinates center of which is coincide with the center of pulsar. Polar axis of this system is perpendicular to the plane of the motion of the center of mass. Then, the unit vecton of the rotation axe $n_p^{(\alpha)}$ has the following vierbein components
\begin{equation}
n_p^{(\alpha)}=\{\sin(\theta_p)\cos(\phi_p-\phi_{\omega}(\tau)),\quad\sin(\theta_p)\sin(\phi_p-\phi_{\omega}(\tau)),\quad\cos{\theta_p}\}.
\end{equation}
Where $\phi_{\omega}(\tau)=\int\limits^{\tau}_0\omega(r(\tau'))d\tau'$ is the rotation angle of the vierbeins, $\theta_p$ and $\phi_p$ are initial angles of a spherical coordinates of the vector $n_p^{(\alpha)}$.

Pulse can be received by observer, if and only if the unit vector $n_{(\alpha)}$ lies between two cones
\begin{equation}\label{extinctions}
\frac{\sqrt{1-\left(n_{(\beta)}n_p^{(\beta)}\right)^2}}{\tan(\alpha_2/2)}<n_{(\beta)}n_p^{(\beta)}<\frac{\sqrt{1-\left(n_{(\beta)}n_p^{(\beta)}\right)^2}}{\tan(\alpha_2/2)},
\end{equation}
where $\alpha_2$ and $\alpha_1$ are the cone angles, that bound the cones of pulsar beam (see, e. g., \cite{Stovall2012}). The results of calculations of the time of observations of the pulsar for some parameters of its orientation is presented in Figure \ref{extinctions}. If the parameters of the motion of binary pulsar are known, by fitting this diagram one can extract the unknown parameters $(\theta_p, \phi_p, \alpha_1, \alpha_2),$ and to use this to predict time of observation of this pulsar in future.
\begin{table}
\begin{ruledtabular}
\caption{Parameters of the puelsar}\label{parametersofpulsar}
\begin{tabular}{l c}
   Parameter & Model value\\
\hline
Initial azimuthal angle of the pulsar axis, $\theta_p$ & 2 rad\\
Initial polar angle of the pulsar axis, $\phi_p$ & 0 rad \\
Conical angle of the inner boundary of the pulsar beam, $\alpha_1$ &  1,35 rad \\
Conical angle of the other boundary of the pulsar beam, $\alpha_2$ &  1,5 rad \\
 \end{tabular}
\end{ruledtabular}
\end{table}
\begin{figure}
\includegraphics[width=0.5\linewidth]{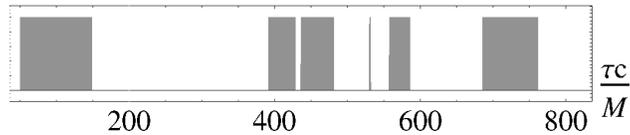}
\caption{The shaded regions represents the proper time in which observation of the pulsar beam is possible. The set of parameters of motion is given in Table \ref{parametersofmotion}. The parameters of the orientation of the pulsar are summarised in Table \ref{parametersofpulsar}.}\label{extinctions}
\end{figure}
The magnifification coefficient of this pulses can be calculated by using the following formula (for definition and derivation see \cite{Tarasenko}):
\begin{equation}\label{magnification_coefficient}
\frac{I_p}{I_0}\equiv K=\frac{1}{r^2\sin\phi_r}\frac{D}{(z+1)^4\sqrt{1-(1-2M/r)D^2/r^2}}\left|\frac{d D}{d\phi_r}\right|.
\end{equation}
Light from a source can be received by observer from infinity number of trajectories (see, e. g., \cite{Tarasenko, BisnovatyiRays}). But the largest intensity has the main ray of light that is received which minimal angle $\phi.$ The intensity of other rays smaller as $\sim e^{-2\pi n},$ where $n$ is number of ray \cite{BisnovatyiRays}.  Because of this, the most intensity rays comes to the observer from the main trajectory. This gives us approach to distinct the lights from this ray from another. As an example on Fig. \ref{MagnificationFactor} we plot the magnification factor (\ref{magnification_coefficient}) for two first rays for the set of parameters of motion \ref{parametersofmotion}.
\begin{figure}
\includegraphics[width=0.7\linewidth]{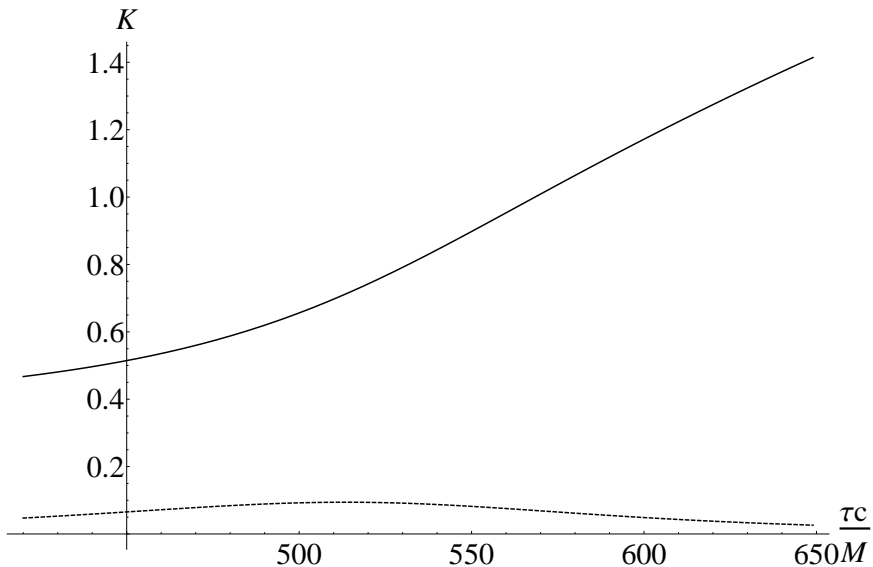}
\caption{Magnification as a function of proper time of the source for the set of parameters in table \ref{parametersofmotion}. Shaded line represents the magnification factor for the main ray, while the dashed line --- for firs order ray (n=1, see also \cite{Tarasenko})}\label{MagnificationFactor}
\end{figure}
\newpage
\section{Discussion}\label{discussions}

In the present work a technique for modeling electromagnetic radiation of a binary system in the field of a supermassive black hole has been presented. We have calculated redshift, magnification coefficient and extinction time for a model binary system. The results suggest that for a sufficiently close binary system the redshift can be of the order of unity and a weak field approximation may not be used.

Redshift of a compact binary has two components: a slowly changing one and a rapidly changing but small one. They are connected with the motion of the system as a whole around the black hole and with the relative motion of stars in the system, respectively. Using the redshift data, the timescales and amplitudes of both components can be estimated, which provides constraints on the orbital parameters of the binary system.

 The part of the redshift that is correspondence to the motion of the system as a whole is more interesting because of this motion can be relativistic. It is follows from numerical calculations that changes in different parameters of the motion leads to characteristic changes of the function of redshift (see Fig. \ref{comparison}). This gives one possibilities to reconstruct of the motion of the binary system by using the redshift as a function of time that is obtained from observations.
\begin{figure}[h]\label{comparison}
\includegraphics[width=0.7\linewidth]{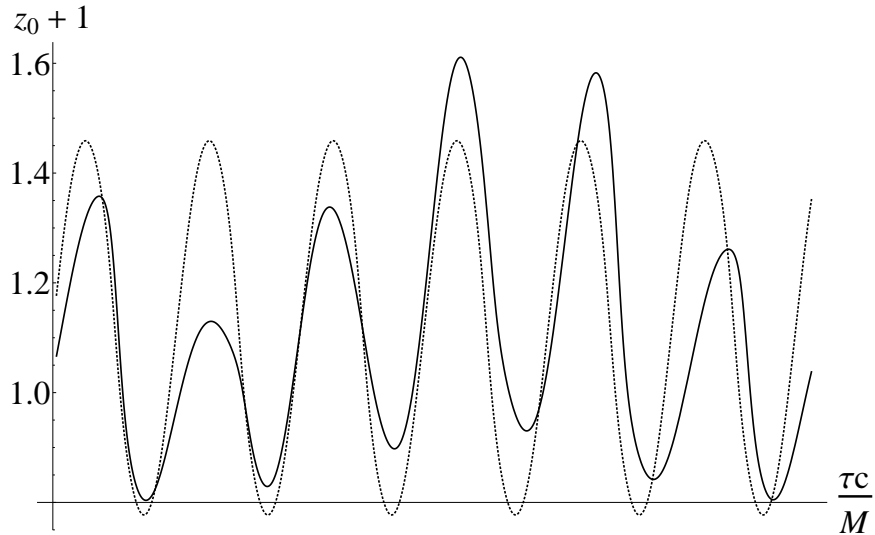}
\caption{Redshift from the center of mass $z_0$ as a function of proper time: dashed --- for the cases of circular motion (with $s_1=s_2=15M$) and solid --- for an extended orbit ($s_1=10M,$ $s_2=40M$). The time axis of this graphics are not in the same scale}\label{comparison}
\end{figure}

The equations of motion of binary star that have been presented in this work can be applied to the cases of all external gravitational fields that changes on a scale much more than the size of the compact binary system. Due to this it is possible to apply the method that is described to the cases of other external gravitational fields.

The using of the Fermi coordinates formalism gives us possibilities to simply derive the condition of times of the extinctions of the pulsar (\ref{extinctions}). In many works the Lorentz transformations approach for this purpose have been used (see  \cite{Wang2009, Stovall2012}.). But this approach do not consists the geodesic precision in the field of supermassive black hole and do not can be used for the calculating of the times of extinction for the pulsar motion in the small vicinity of black hole.

For practical applications it is much more interesting to solve the full inverse problem: given the redshift as a function of time, find orbital parameters of the binary system. In some works the inverse problem for the source that moves in the field of supermassive black hole has been solved \cite{Tarasenko} by using some additional data, such us magnification factor of the ray. It is interesting to solve inverse problem for a binary star in external gravitational field by using the redshift data only, that can be obtained with high accuracy (the other characteristic of electromagnetic radiation such us magnification coefficient are obtained with much less accuracy). We leave it for a separate paper.

A decomposition of redshift of a compact radiation source into a series has been obtained. This expansion can be useful not only for a binary stars, but for any compact source in external gravitational field for which the law of internal motion is obtained separately in a comoving reference frame.

\appendix
\section{Wave vector properties}

Consider a radiation source moving along a trajectory $x^i=x^i(\tau)$, and an observer staying at the point $x=(r=\infty,\theta_O,\varphi_O)$ at the infinity. Let $k^i(x^l)$ be a wave vector of the light ray the was emitted by the source at the point $x^i$ and will be received by the observer. Using equation of geodesic, $k^i(x^l)$ can be obtained for every point of the spacetime (in the case when there are several such rays, a concrete one is considered, so that $k^i$ is a smooth function of coordinates).

Vector $k^i$ is isotropic and satisfies equation of geodesic:
\begin{equation}\label{1}
    k^i k_i = 0
\end{equation}
\begin{equation}\label{2}
    k^l k_{i;l} = 0
\end{equation}

Let $\xi^i$ be a Killing vector ($\xi_{i;k}+\xi_{k;i}=0$). Equations (\ref{1}) and (\ref{2}) imply that
\begin{equation}
    k^i (\xi^l k_l)_{,i} = 0,
\end{equation}
which means that $\xi^l k_l$ is constant along the ray. If the space-time is static $\dfrac{\partial}{\partial t}$ is Killing vector and we have $k_4=A=const.$ By using the appropriate parametrization of the isotropic geodesic, we can establish that $A=1$ and therefore $k_4=1$ in whole space time.

For a static spherically-symmetric spacetime $k_i$ is satisfied the following relations:
\begin{equation}\label{5}
    k_{i;j}-k_{j;i}=0.
\end{equation}
To prove this, introduce spherical coordinates $x^i=(r,\theta,\varphi,ct)$ so that the observer is located at the pole $\theta=0$. In this case $k_i$ does not depend on time and angle $\varphi$. Trajectory of each ray lies in a plane for which $\varphi=const$, hence $k^3=0$. $k_4(x)=1$ by it definition.  Equations (\ref{1}) and (\ref{2}) imply that
\begin{equation}
    k^l(k_{i,l}-k_{l,i})=0,
\end{equation}
which leads to
\begin{equation}\label{12}
\begin{split}
    &k_{1,2}-k_{2,1}=0;\\
    &k_{3,j}=0;\\
    &k_{4,j}=0.
\end{split}
\end{equation}
The covariant form of the relations  (\ref{12}) gives (\ref{5}), which completes the proof.

\section{Redshift of a finite-size radiation source}
For a finite-size source equation (\ref{redshiftfinal}) can be applied to each part of the source. In this case $k^i$ is calculated at the location of a corresponding part of the source. For a compact source the redsift can be expanded into a series using $\rho/r$ as a parameter, where $\rho$ is a characteristic size of the source and $r$ is the characteristic distance, i. e. distance to the field center.

Let introduce some inner "point" of the body of the source $C$ with coordinates $x^i_C$ that moves along world line $x^i_C=\xi^i(\tau)$ and the considering part be located at a point $P$ with coordinates $x^i_P$. Our aim is to express the redshift of emitted rays from the point $P$ in terms of quantities that are defined at the world line $\xi^i(\tau)$. Consider a geodesic $x^i(\sigma)$ that is orthogonal to the world line $\xi^i(\tau)$:
\begin{equation}\label{10}
    \frac{d^2x^i}{d\sigma^2}+\Gamma^i_{kl}\frac{dx^k}{d\sigma}\frac{dx^l}{d\sigma}=0,
\end{equation}
\begin{equation}\label{12a}
u_i\frac{dx^i}{d\sigma}\left(x^j_C\right)=0,
\end{equation}
where $\Gamma^i_{kl}$ are the Christoffel symbols, $u^i=d\xi^i/d\tau$, $\sigma$ is the parameter that is equal to the geodesic distance. Denote
\begin{equation}\label{13}
\eta^i=\frac{dx^i}{d\sigma}\left(x^j_C\right).
\end{equation}
Equations (\ref{10})-(\ref{13}) allow to express $x^i_P$ using $\eta^i$:
\begin{equation}\label{14}
x^i_P=\xi^i(\tau)+\eta^i\sigma_P+O(\rho^2).
\end{equation}
Where $\sigma_P$ denotes the geodesic distance from $x^i_C$ to $x^i_P.$ This has the order of $\rho.$
The vectors $k^i\left(x^i_P\right)$ and $u^i\left(x^i_P\right)$ that are included in (\ref{redshiftfinal}) must be calculated on the world line of the source. To find the corresponding quantities on the $\xi^i(\tau),$ it is necessary to introduce the vector fields $k^i_F(x^j)$ and $u^i_F(x^j)$ on the interval between $x^i_C$ and $x^i_P,$ by translating this vectors covariantly parallel along the geodesic that connect the points $x^i_C, x^i_P.$ Therefore we obtain
\begin{equation}\label{fields}
\begin{split}
&k^i_F(P)=k^i(P);\qquad u^i_F(P)=u^i(P).\\
&k^i_F(C)=k^i_F(P)+\Gamma^i{}_{sl}k^i_F(P)\eta^l\sigma_P+O(\rho^2),\\
&u^i_F(C)=u^i_F(P)+\Gamma^i{}_{sl}u^i_F(P)\eta^l\sigma_P+O(\rho^2).
\end{split}
\end{equation}
Apart from the fields that have been introduced we have another vector field $k^i(x^j).$ This field can be determined as the field of all tangent vectors to the null geodesics (that are 1 order) that are leaves to the observer (see Appendix A). We also have $k^i(P)=k^i_F(P)$
\begin{equation}\label{k^i_D(x)}
k^i_F(P)=k^i(P)=k^i(C)+k^i{}_{,j}(C)h^j{}_{(\alpha)}X_1^{(\alpha)}+O(\rho^2).
\end{equation}
Where $X_1^{(\alpha)}$ --- is the Fermi coordinates of the source $(P)$ relative to the center of mass in $(C)$.
By substituting (\ref{k^i_D(x)}) into (\ref{fields}), we have
\begin{equation}\label{k^i_F}
k^i_F(C)=k^i(C)+k^i{}_{;j}(C)h^j{}_{(\alpha)}X_1^{(\alpha)}+O(\rho^2).
\end{equation}
From (\ref{14}) and the equation $\sigma_P\eta^i=h^i_{(\alpha)}X^{(\alpha)}$ for the 4-velocity of the source we have
\begin{equation}
\begin{split}
&u^i(P)=\frac{dx^i_P}{d\tau}=\frac{dx^i_C}{d\tau}+\frac{d}{d\tau}\left(h^i{}_{(\alpha)}X_1^{(\alpha)}\right)=\\
&h^i{}_{(4)}+h^i{}_{(\alpha)}v^{(\alpha)}+e_{(\alpha)}{}^{(\beta)(\gamma)}h^i{}_{(\beta)}\omega_{(\gamma)}X_1^{(\alpha)}-\Gamma^i{}_{sl}h^s{}_{(\alpha)}X_1^{(\alpha)}h^l{}_{(4)}.
\end{split}
\end{equation}
Then
\begin{equation}\label{u^i_F}
u^i_F(C)=h^i{}_{(4)}+h^i{}_{(\alpha)}v^{(\alpha)}+e_{(\alpha)}{}^{(\beta)(\gamma)}h^i{}_{(\beta)}\omega_{(\gamma)}X_1^{(\alpha)}.
\end{equation}
Since the fields $k^i_F(x^j),$ $u^i_F(x^j)$ and $g_{ij}(x^j)$ are covariantly constant along the geodesic from $x^i_P$ to $x^i_C$ we have
\begin{equation}\label{transfer}
\left(k^iu_i\right)_s=g_{ij}(P)k^i_F(P)u^j_F(P)=g_{ij}(C)k^i_F(C)u^j_F(C).
\end{equation}
By substituting (\ref{k^i_F}) and (\ref{u^i_F}) into (\ref{transfer}) with metric in the form $g_{ij}(C)=g_{ij}{}^{Sch}+\phi_{(l)(m)}h^{(l)}{}_i h^{(m)}{}_j$ (see also (\ref{Schwarzschild})) and (\ref{redshiftfinal}) we obtain
\begin{equation}\label{redshiftdecomposition}
\begin{split}
&z_{\infty}(\tau+\tau_{ret})+1=h_{(4)i}k^i+v_{(\alpha)}k^{(\alpha)}+e_{(\alpha)(\beta)(\gamma)}k^{(\alpha)} X_1^{(\gamma)}\omega^{(\beta)}+h_{(4)i}k^i{}_{;j}h^{j}{}_{(\alpha)}X_1^{(\alpha)}+2\phi_{(4)(4)}k^{(4)}\\
&+O(\rho^2,v^2,\rho v,\phi\rho,\phi v).
\end{split}
\end{equation}
Where $k^{(l)}=h^{(l)}{}_ik^i.$
All terms in (\ref{redshiftdecomposition}) must be calculated on the world line $\xi(\tau)$ at the proper time $\tau.$ The term $\phi_{(4)(4)}k^{(4)}$ in this expression has the sense of Shapiro delay due to the gravitational field of each pulsar. It can be found as
\begin{equation}
\begin{split}
&2\phi_{(4)(4)}k^{(4)}=\Delta_1+\Delta_2,\quad{where}\\
&\Delta_1=4\frac{m_1}{R_0}(z_0+1),\quad \Delta_2=\frac{m_2}{\sqrt{x^{(\alpha)}x_{(\alpha)}}}(z_0+1).
\end{split}
\end{equation}
The $\Delta_2$ has the order of $\Delta_2\sim m/\rho.$ And usually this term $\lesssim 10^{-6}$ for a pulsar and therefore we neglect of this term. The another therm  $\Delta_1$  has the form $\Delta_1=const\cdot (z_0+1).$ For a pulsar this leads only to rescaling of the quantity $T_p$ (see Sec. \ref{redshift}) and therefore we will not consider of this term.

The time dilation due to the finite size of the compact system we denote as $\tau_{ret}$. This time dilation can be found from the relation $k_i\dfrac{dx^i}{d\lambda}=const$ that is hold along isotropic geodesic (see e. g. \cite{Sing}). Here $\dfrac{dx^i}{d\lambda}$ is a vector between two close geodesics in sheaf, and $\lambda$ is a global parameter that numbers geodesics in this sheaf. We obtain
\begin{equation}\label{timedilation}
\begin{split}
&(k_iu^i)_o(z+1)\tau_{ret}=(k_i\eta^i)_s\sigma_P+O(\rho^2)\Rightarrow\\
&\tau_{ret}=\frac{1}{A(z+1)}k_{(\alpha)}(C)X_1^{(\alpha)}+O(\rho^2).
\end{split}
\end{equation}
By substituting (\ref{timedilation}) into (\ref{redshiftdecomposition}) we obtain
\begin{equation}\label{zrepresentation}
\begin{split}
&z_{\infty}(\tau)+1=h_{(4)i}k^i-\frac{1}{z_0+1}\frac{dz_0}{d\tau}k_{(\alpha)} X_1^{(\alpha)}+v_{(\alpha)}k^{(\alpha)}+e_{(\alpha)(\beta)(\gamma)}k^{(\alpha)} X_1^{(\gamma)}\omega^{(\beta)}+h_{(4)i}k^i_{;j}h^{j}{}_{(\alpha)}X_1^{(\alpha)}\\
&+O(\rho^2,v^2,\rho v).
\end{split}
\end{equation}
Where we denote $z_0$ from the relation $z_0+1=k_i\dfrac{dx^i_C}{d\tau}$.
On taking into account the relations $k_{i;j}=k_{j;i},$ (see Appendix A) we can rewrite (\ref{zrepresentation}) as
\begin{equation}\label{zz1}
z_{\infty}(\tau)+1=(z_0+1)\left(1-\frac{d}{d\tau}(n_{(\alpha)}X_1^{(\alpha)})\right)+O(\rho^2,v^2,\rho v).
\end{equation}
Where $n^{(\alpha)}=\dfrac{k^{(\alpha)}}{\sqrt{k_{(\alpha)}k^{(\alpha)}}}=\dfrac{k^{(\alpha)}}{(z_0+1)}$ --- is the unit vector of the ray. All terms in right-hand side of equation (\ref{zz1}) must be calculated at the proper time $\tau.$

Note that the formula (\ref{zz1}) can be rewritten in covariant form:
\begin{equation}
z_{\infty}(\tau)+1=(z_0+1)\left(1-\frac{d}{d\tau}(n_i\eta^i\sigma_p))\right)+O(\rho^2,v^2,\rho v).
\end{equation}
Where $\eta^i\sigma_p\approx\delta x^i$ is the coordinate distance between the source and the center of mass at the same proper time, $n_i=\dfrac{k_i}{(z_0+1)}.$
\bibliography{inverse02_rev}

\end{document}